\documentclass[12pt,preprint]{aastex}

\usepackage{emulateapj5}

\begin{document}

\title{A class of self-gravitating, magnetized accretion disks}

\author{Mohsen Shadmehri and Fazeleh Khajenabi
\affil{Department of Physics, School of Science, Ferdowsi
University, Mashhad,
Iran}}\email{mshadmehri@science1.um.ac.ir,fkhajenabi@science1.um.ac.ir}

\begin{abstract}
The steady-state structure of self-gravitating, magnetized accretion
disks is studied using a set of self-similar solutions which are
appropriate in the outer regions. The disk is assumed to be
isothermal  and the magnetic field outside of the disk is treated
in a phenomenological way. However, the internal field is determined
self-consistently. The behaviour of the solutions are
investigated by changing the input parameters of the model, i.e.
mass accretion rate, coefficients of  viscosity and 
resistivity, and the magnetic field configuration.
\end{abstract}


\section{Introduction}
One of the key physical ingredients in accretion disks is
self-gravity. It  plays a significant role in many
such systems, ranging from protostellar disks to Active
Galactic Nuclei (AGN). The radial and  vertical equations for
the disk structure are significantly modified because of the possible
impact of the self-gravity, although traditional models of
accretion disks ignore self-gravity just for
simplicity (e.g., Pringle 1981). Nevertheless, Deviations from Keplerian
rotation in some AGN and the flat infrared spectrum of some T Tau
stars can both  be described by self-gravitating disk models.

Because of the complexity of the equations and in order to obtain
analytical  results, authors have  studied the effects
related to disk self-gravity either in the vertical structure of
the disk (e.g., Bardou et al. 1998) or in the radial direction
(e.g., Bodo \& Curir 1992). In a situation where we lack strong
empirical evidence for the detailed mechanisms involved, Bertin (1997) proposed
a new class of self-gravitating disks for which efficient cooling
mechanisms are assumed to operate so that the disk is
self-regulated at a condition of approximate marginal Jeans
stability. Thus, the energy equation is replaced by a
self-regulation prescription. He showed that in the absence of a
central point mass, there is a set of self-similar solutions
describing the steady-state structure of such self-regulated
accretion disks. The self-similar solution corresponds to a flat rotation
curve, while the disk has a fixed opening angle. In fact, Bertin's self-similar
solution is a generalization of the self-similar solution for 
self-gravitating disks (Mestel 1963) dominated by viscosity.

Subsequent analysis confirmed the validity of this simple self-similar
solution (Bertin \& Lodato 1999). Moreover, such kinds of
self-regulated accretion disk can successfully describe the
spectral energy distribution of protostellar disks (Lodato \&
Bertin 2001). Recently, Lodato \& Rice (2004) studied the
transport associated with gravitational instabilities in a
relatively cold disk using numerical simulations. They showed
that the disk truly settles into a self-regulated state, where the
axisymmetric stability parameter $Q\approx 1$ and where transport
and energy dissipation are dominated by self-gravity. However,
these studies neglected the possible effects of magnetic fields. Many
authors tried to construct models for magnetized disk either in a
phenomenological way based on some physical considerations (e.g.,
Schwartzman 1971; Lovelace et al. 1994; Kaburaki 2000; Shadmehri
2004; Shadmehri \& Khajenabi 2005) or by direct numerical
simulations (Stone et al. 1999; McKinney \& Gammie 2002).
Bisnovatyi-Kogan \& Lovelace (2000) suggested that recent papers
discussing Advection-Dominated Accretion Flow (ADAF) as a possible solution for astrophysical
accretion should be treated with caution, particularly because of our
ignorance surrounding the magnetic field. Models of magnetized accretion
disks with externally imposed large scale vertical magnetic field
and anomalous magnetic field due to enhanced turbulent diffusion
have also been studied (see, e.g., Campbell 2000; Ogilvie \&
Livio 2001). These models are restricted to subsonic turbulence
in the disk and the viscosity and magnetic diffusivity are due to
hydrodynamic turbulence.

Recently, we presented a set of analytical self-similar solutions
for the steady-state structure of a magnetized,
radiation-dominated disk (Shadmehri \& Khajenabi 2005). Although
this analysis is just a simple model, one can see some possible
effects of the magnetic field on the structure of
radiation-dominated disks, at least at a fundamental level.
However, we applied some simplifying assumptions 
concerning  the magnetic field both outside and inside of the disk.
Our approach was based on interesting studies by Lovelace et al.
(1987) and Lovelace et al. (1994). We noticed that it is possible to
take a similar approach to study magnetized, self-gravitating
disk.

In this paper, we relax the self-regulation condition  of Bertin
(1997) simply by assuming an isothermal equation of state. We
present self-similar solution for the steady-state structure of
such self-gravitating disk for which the effect of the magnetic field
is taken into account. While the magnetic field outside of the
disk is studied in a phenomenological approach similar to
Lovelace et al. (1994), the magnetic field inside  the disk is
obtained self-consistently. We show that magnetic fields  may
cause significant changes in the typical behaviours of the solutions
compared to the nonmagnetized case. The basic equations of the
model are described in section 2. Self-similar solutions are
obtained in section 3. In spite of the similarity of Bertin's
solutions (1997) to ours, there are differences between his work
and the present analysis. A summary of results are discussed in
the final section.

\section{General Formulation}
The basic equations are integrated over the vertical thickness of
the disk. The mass continuity equation becomes
\begin{equation}
-2\pi R \Sigma v_{\rm R}=\dot{M},\label{eq:masscon}
\end{equation}
where $v_{\rm R}$ is the radial velocity and $\Sigma=\int dz \rho
\simeq 2 h \rho $ is the surface density of the disk. The
half-thickness is denoted by $h$, where we consider the magnetic
field effect on the disk thickness. We will see that the disk can
be compressed or flattened depending on the field configuration.
Also, we note that since the radial velocity is negative for
accretion (i.e., $v_{\rm R}<0$), the accretion rate $\dot{M}$ as
an input parameter of our model is positive.

The radial momentum equation reads
\begin{equation}
\Sigma v_{\rm R}\frac{d v_{\rm R}}{d
R}-\Sigma\frac{v_{\varphi}^2}{R}=-\frac{d P}{d
R}-\Sigma\frac{GM(R)}{R^2}+\int F_{\rm R}^{\rm mag}
dz,\label{eq:rcom}
\end{equation}
where $P=\int dz p$ is the integrated disk pressure and $M(R)$ is
the mass of the disk within radius $R$ . We are neglecting the
mass of the central object comparing to the disk mass. This is relevant to 
protostellar disks at the beginning of the accretion phase during
which the mass of the central object is small and self-gravity of
the disk plays an important role. Also, one may argue that our
model corresponds to disks at large radii because the effects of
the central mass become unimportant in the outer regions of the
disk. The term associated with self-gravity of the disk, i.e.
$GM(R)/R^{2}$, is generally applicable to spherical geometry. In
fact, the correct term takes on an integral form (see equation
(4) of Bertin 1997). But, as we will show, our solution describes a
disk in which the surface density is inversely proportional to
radius. Interestingly, for such a disk which resembles a Mestel
disk, the correct gravitational force is known which is exactly
equal to our approximate formula (Binney \& Tremaine 1987). It
implies that  we  consider the exact gravity force  in
our equations.  Now, we can write
\begin{equation}
\frac{dM(R)}{dR}=2\pi R\Sigma. \label{eq:gravity}
\end{equation}

The last term of equation (\ref{eq:rcom}) represents the height-integrated radial magnetic force which can be written as
(Lovelace et al. 1994; hereafter LRN)
\begin{displaymath}
\int F_{\rm R}^{\rm mag} dz=\frac{1}{2\pi}(B_{\rm R} B_{\rm
z})_{\rm h}-\frac{1}{4\pi R^2}\frac{d}{d R}[h
R^{2}<B_{\varphi}^{2}-B_{\rm R}^2>]
\end{displaymath}
\begin{equation}
-\frac{1}{4\pi}\frac{d}{d R}[h <B_{\rm
z}^{2}>]+\frac{1}{4\pi}\frac{dh}{dR}(B_{\varphi}^{2}+B_{\rm z
}^{2}-B_{\rm R}^{2})_{\rm h},
\end{equation}
where $<\cdots>\equiv\int_{-h}^{h}dz(\cdots)/(2h)$, and the $h$
subscript denotes that the quantity is evaluated at the upper
disk plane, i.e. $z=h$. Similarly, integration over $z$ of the
azimuthal  equation of motion gives

\begin{equation}
R\Sigma v_{\rm R}\frac{d}{d R}(Rv_{\varphi})=\frac{d}{d
R}[R^{3}\nu\Sigma\frac{d}{d R}(\frac{v_{\varphi}}{R})]+\int
R^{2}F_{\varphi}^{\rm mag} dz,\label{eq:phicom}
\end{equation}
where
\begin{displaymath}
\int R^{2}F_{\varphi}^{\rm mag}
dz=\frac{1}{2\pi}(R^{2}B_{\varphi}B_{\rm z})_{\rm
h}-\frac{1}{2\pi}\frac{dh}{dR} (R^{2}B_{\rm R}B_{\varphi})_{\rm h}
\end{displaymath}
\begin{equation}
+\frac{1}{2\pi}\frac{d}{dR}[hR^{2} <B_{\rm R}B_{\varphi}>].
\end{equation}
Here, the last term of equation (\ref{eq:phicom}) represents the
height-integrated toroidal component of magnetic force multiplied
by $R^{2}$.

In our model, we also assume (Shakura \& Sunyaev 1973)
\begin{equation}
\nu=\alpha c_{\rm s} h,\label{eq:vis}
\end{equation}
where $c_{\rm s}$ is the local sound speed and $\alpha$ is a
constant less than unity.

The $z$ component of equation of motion gives the condition for
vertical hydrostatic balance, which can be written as
\begin{equation}
\frac{\Sigma}{h}c_{\rm s}^{2}=\pi
G\Sigma^{2}+(\frac{1}{4\pi})[(B_{\rm R})_{\rm
h}^{2}+(B_{\varphi})_{\rm h }^{2}]-\frac{h}{4\pi}(B_{\rm R})_{\rm
h}\frac{dB_{\rm z}}{dR},\label{eq:zcom}
\end{equation}
where the term on the left hand side of the above equation
corresponds to the thermal pressure. Since we neglect the central
object, the self-gravity of the disk in the vertical direction
leads to the first term on the right hand side of the above
equation.  Now, we can treat the internal magnetic field using the
induction equation. LRN showed that the variation of $B_{\rm z}$
with $z$ within the disk is negligible for even field symmetry.
Moreover, $B_{\rm R}$ and $B_{\varphi}$ are odd functions of $z$
and consequently $\partial B_{\rm R}/\partial z\approx (B_{\rm
R})_{\rm h}/h$ and $\partial B_{\varphi}/\partial z\approx
(B_{\varphi})_{\rm h}/h$. Krasnopolsky \& Konigl (2002) applied
similar configurations to study the time-dependent collapse of
magnetized, self-gravitating disks. Thus,
\begin{displaymath}
B_{\rm R}(R,z)=\frac{z}{h}(B_{\rm R})_{\rm h},
B_{\varphi}(R,z)=\frac{z}{h}(B_{\varphi})_{\rm h},
\end{displaymath}
\begin{equation}
 B_{\rm z}(R,z)=B_{\rm z}(R),
\end{equation}
and the induction equation reads
\begin{equation}
-RB_{\rm z} v_{\rm R}-\frac{\eta R}{h}(B_{\rm R})_{\rm h}+\eta
R\frac{d B_{\rm z}}{d R}=0,\label{eq:induction}
\end{equation}
where the magnetic diffusivity $\eta$ has the same units as
kinematic viscosity. We assume that the magnitude of $\eta$ is
comparable to that of the turbulent viscosity $\nu$ (e.g.,
Bisnovatyi-Kogan \& Ruzmaikin 1976; Shadmehri 2004). Exactly in
analogy to the alpha prescription for $\nu$, we are using a similar
form for the magnetic diffusivity $\eta$,
\begin{equation}
\eta=\eta_{0}c_{\rm s}h,
\end{equation}
where $\eta_{0}$ is a constant. Note that $\eta$ is {\it not}
constant and depends on the physical variables of the flow, and
in our self-similar solutions, as we will show, $\eta$ scales
with radius as a power law. This form of scaling for diffusivity
has been widely used by many authors (e.g., Lovelace, Wang \&
Sulkanen 1987; Lovelace, Romanova \& Newman 1994; Ogilvie \&
Livio 2001; R\"{u}diger \& Shalybkov 2002).

While equation (\ref{eq:induction}) describes the transport of a
large-scale magnetic field (here, $B_{\rm z}(R)$), the values of
$(B_{\rm R})_{\rm h}$ and $(B_{\varphi})_{\rm h}$ are determined
by the field solutions external to the disk. Instead, we are
following the approach of LNR, in which the external field solutions
obey the relations
\begin{equation}
(B_{\rm R})_{\rm h}=\beta_{\rm r} B_{\rm z}, (B_{\varphi})_{\rm
h}=\beta_{\varphi} B_{\rm z},
\end{equation}
where $\beta_{\rm r}$ and $\beta_{\varphi}$ are constants of
order unity ($\beta_{\varphi}<0$). Thus, one can simply show that
$<B_{\rm R}^{2}>=\beta_{\rm r}^{2} B_{\rm z}^{2}/3$,
$<B_{\varphi}^{2}>=\beta_{\varphi}^{2} B_{\rm z}^{2}/3$ and
$<B_{\rm R}B_{\varphi}>=\beta_{\rm r}\beta_{\varphi}B_{\rm
z}^{2}/3$. Note that these beta factors are really empirical
scalings which now have abundant support from computer
simulations of MHD and Poynting outflows from disks. Ustyugova et
al. (1999) gives a  detailed analysis and provides  strong
evidence for $\beta_{\rm r} \sim |\beta_{\varphi}| \sim 1$.

To close the equations of our model, we can write the
self-regulation condition as has been done by Bertin (1997). But
one should naturally expect another form of self-regulation
prescription in the magnetized case. In fact, self-regulation
results from a competition of dissipation and instabilities.
Indeed, the instability of a magnetized disk is different from
that of unmagnetized disk. Effects of the magnetic field on linear
gravitational instabilities in two-dimensional differentially
rotating disks have been investigated in detail by Elmegreen
(1987, 1994), Gammie (1996) and Fan \& Lou (1997). Axisymmetric
and nonaxisymmetric perturbations show significantly different
behaviours. While axisymmetric instability in a thin disk borrows
its physical grounds from Toomre (1964) and requires $Q<1$, the
literature has lacked an explicit theoretical evaluation of
critical $Q$ for nonaxisymmetric gravitational runaways. In order
to avoid such difficulties one can consider the axisymmetric
stability. The presence of the magnetic fields modifies the Toomre
criterion for a hydrodynamic disk in such a way that magnetized
disks are unstable to axisymmetric perturbation if $Q_{\rm M}=Q
(1+1/\beta)<1$, where $\beta$ is the ratio of the thermal
pressure to the magnetic pressure (e.g., Shu 1992).

There is a fundamental point concerning the self-regulation
prescription. When the disk mass becomes large enough to
induce {\it global} instability there is nothing like
self-regulation taking place. We can see numerical simulations
which describe massive unstable disk (e.g., Bonnell 1994;
Matsumoto \& Hanawa 2003). However, considering existing numerical
simulations, we think, it is not possible simply to say that all
massive disks fragment because, to our knowledge, not only current
numerical simulations suffer from their own limitations, but
also they do not generally give a fully consistent  picture as
for global fragmentation. This situation becomes more complicated 
when we consider magnetic fields. It seems that there is a complicated
interaction between gravitational instability and MHD turbulence
that influences disk structure, but MHD turbulence reduces the
strength of the gravitational instability (Fromang 2005).

Almost all numerical simulations show that global instabilities
are highly depend on the thermal state of the disk (Gammie 2001,
Johnson \& Gammie 2003). However, thermal state of the disk has
been neglected by some authors and so their approach is not a
complete analysis. Simulations by Mastsumoto \& Hanawa (2003)
which do not include detailed thermal evolution, predict
fragmentation in an early phase. But all these fragments are on
tight orbits and are likely to merge due to disk accretion.
Matzner \& Levin (2005) argue analytically that viscous heating
and stellar irradiation quenches fragmentation and conclude that
numerical simulations lead to fragmentation which do not account
for irradiation and  are unrealistic. On the other hand, not
necessarily all numerical simulations of massive disks lead to
fragmentation. For example, Pickett at al. (2000) failed to
obtain fragmentation in similar thermodynamics condition. Thus,
 whether or not massive disks fragment due to global
instability remains controversial. At this stage what we can say
is  that even if a massive disk with no star in the center does
not fragment, it does not mean that self-regulation is occurring
in the disk. In order to avoid such difficulties, we relax the 
self-regulation condition and replace it by an isothermal
assumption, i.e.
\begin{equation}
P=\Sigma c_{\rm s}^{2},\label{eq:iso}
\end{equation}
where $c_{\rm s}$ is constant sound speed. We hope our simple
approach will be able to illustrate some possible effects of the
magnetic fields on the structure of self-gravitating disks.

\section{Self-similar solutions}

Equations (\ref{eq:masscon}), (\ref{eq:rcom}),
(\ref{eq:gravity}), (\ref{eq:phicom}), (\ref{eq:zcom}),
(\ref{eq:induction}) and (\ref{eq:iso})  constitute the basic
equations of our model for the steady-state structure of
self-regulated magnetized disk. We can solve these equations
numerically using appropriate boundary conditions. However,
before doing such an analysis it would be illustrative to study
the typical behaviour of the solutions by applying
semi-analytical methods, e.g. self-similarity. Indeed, any
self-similar solution contains part of the behavior of the system,
in particular far from the boundaries. However the main goal of our
analysis is just to illustrate the possible effects of the magnetic
field on the steady-state structure of a self-gravitating disk.
For this propose, we think, self-similar solutions are very
useful.

It is then straightforward to find a self-similar solution with
the radial dependence
\begin{equation}
\frac{\Sigma(R)}{\Sigma_{0}}=a(\frac{R}{R_{0}})^{-1},
\end{equation}
\begin{equation}
\frac{v_{\varphi}(R)}{V_{0}}=b ,
\end{equation}
\begin{equation}
\frac{v_{\rm R}(R)}{V_{0}}=-c ,
\end{equation}
\begin{equation}
\frac{P(R)}{P_0}=d(\frac{R}{R_{0}})^{-1},
\end{equation}
\begin{equation}
\frac{B_{\rm z}(R)}{B_0}=e (\frac{R}{R_{0}})^{-1},
\end{equation}
\begin{equation}
\frac{h(R)}{R_{0}}=f(\frac{R}{R_{0}}),
\end{equation}
\begin{equation}
\frac{M(R)}{M_0}=q(\frac{R}{R_{0}}),
\end{equation}
with $a$, $b$, $c$, $d$, $e$, $f$ and $q$ being numerical
constants which can be obtained from the equations (see below).
Also,  $R_{0}$, $\Sigma_0$, $P_0$, $V_0$ and $B_0$ are convenient
units which reduce the equations into non-dimensional forms. We are
assuming $V_{0}=c_{\rm s}=\sqrt{P_{0}/\Sigma_{0}}$ and
$M_{0}=R_{0}^{2}\Sigma_{0}$.
If we substitute the above self-similar solutions in the main
equations of the model,  the following system of dimensionless
algebraic equations are obtained which are to be solved for $a$, $b$,
$c$, $d$, $e$, and $f$:
\begin{equation}
ac=\dot{m},
\end{equation}
\begin{equation}
-(ab^{2}+d)=-2a^{2}+[2\beta_{\rm
r}+\frac{2}{3}f(3+\beta_{\varphi}^{2}-\beta_{\rm  r}^{2})] e^{2},
\end{equation}
\begin{equation}
ab(\alpha
f\sqrt{\frac{d}{a}}-c)=\beta_{\varphi}(2-\frac{4}{3}f\beta_{\rm r
})e^{2},
\end{equation}
\begin{equation}
\frac{d}{f}=a^{2}+(f\beta_{\rm r}+\beta_{\rm
r}^{2}+\beta_{\varphi}^{2})e^{2},
\end{equation}
\begin{equation}
c-\eta_{0}\beta_{\rm
r}\sqrt{\frac{d}{a}}-\eta_{0}\sqrt{\frac{d}{a}}f=0,
\end{equation}
\begin{equation}
d = a,
\end{equation}
where $\dot{m}$, $\alpha$, $\eta_0$, $\beta_{\rm r}$ and
$\beta_{\varphi}$ are the input parameters.  Also,
$\dot{m}=\dot{M}/(2\pi R_{0}\Sigma_{0}V_{0})$ is the nondimensional
mass accretion rate. The radial  dependence of physical
quantities both in the magnetized and the unmagnetized cases are
similar. After some algebraic manipulations we can find a
complicated algebraic equation for $f$ and clearly only the real
root is a physical solution. Having the real root of  this
equation, the other physical quantities are determined using the
above equations.

Now, we explore the parameter space of the input parameters and their
effects on the solutions. We restrict our study to values
around unity for $\beta_{\rm r}$ and $\beta_{\varphi}$ (Ustyugova
et al. 1999). Since we find that the sensitivity of solutions on the
parameter $\beta_{\varphi}$ is not very strong, we plot all the
physical variables as a function of $\beta_{\rm r}$ for a fixed
$\beta_{\varphi}$. However, we comment about possible effects of
variations of $\beta_{\varphi}$ by the final figure.

Figure 1 shows various physical variables as function of
$\beta_{\rm r}$ with $\alpha=0.1$, $\eta_{0}=0.1$,
$\beta_{\varphi}=-0.8$ and the mass accretion rate is indicated
on the plots. This figure helps us to see the dependence of the
physical variables on the variations of the mass accretion rate
$\dot{m}$. The plots on the top show surface density (left) and
the opening angle (right) for $\dot{m}=0.5$, $1.0$ and $2.0$. We
see as the parameter of the radial component of the magnetic
field at the surface of the disk increases, the surface density
decreases for a fixed mass accretion rate. But the surface
density increases by increasing $\dot{m}$. Also, as the mass
accretion rate increases, the disk becomes thinner. However, the
disk thickness increases with $\beta_{\rm r}$ for a fixed
accretion rate. The plots on the middle show typical behaviors of
the radial and the rotational velocities. While the ratio of the
radial velocity to the sound speed increases with increasing
$\beta_{\rm r}$,  the ratio of the rotational velocity to the
sound speed decreases keeping $\dot{m}$ constant. However,
increasing the mass accretion rate causes the radial velocity
to decrease. But for the rotational velocity, we see completely
different behaviour, i.e. the ratio $v_{\varphi}/c_{\rm s}$
significantly increases with the accretion rate $\dot{m}$.

Our self-similar solution generally corresponds to $\beta > 1$,
where $\beta$ is the ratio of thermal pressure to magnetic
pressure at the surface of the disk. The plot on the bottom of
Figure 1 (left) shows this behavior, although $\beta$ decreases
with $\beta_{\rm r}$ for a fixed $\dot{m}$. However, if the accretion
rate increases the ratio of the thermal pressure to the magnetic
pressure $\beta$ increases which implies the effect of the
magnetic field on the structure becomes weaker. We have already
discussed about the Toomre parameter $Q$ and a modified version
of this parameter, i.e. $Q_{\rm M}$ because of the magnetic
field. Figure 1 shows variation of both $Q$ and $Q_{\rm M}$ as
function of $\beta_{\rm r}$ (curves corresponding to $Q_{\rm M}$
are shown by dotted lines). Generally, we have $Q_{\rm M}>Q$
because of the dynamical effects of the magnetic field and as
$\beta_{\rm r}$ increases these parameters increases to values
closer to unity or even higher values.  It implies a more stable
disk according to the Toomre criteria as $\beta_{\rm r}$
increases. Also, the Toomre parameter rapidly increases when the
mass accretion rate decreases.

The $\eta_{0}$-dependence of the solutions is shown in the Figure
2. In fact,  this Figure is similar to Figure 1 except for
changing the input parameter $\eta_0$ to different values $0.05$,
$0.1$ and $0.2$. The other input parameter are the same as in  Figure
1. The surface density $\Sigma$, the rotational velocity
$v_{\varphi}$ and the ratio $\beta$  decrease with increasing
$\eta_{0}$. But the radial velocity $v_{\rm R}$ increases with
increasing $\eta_{0}$. We see that as $\eta_0$ tends to the
larger values, the disk thickness increases. Also, for a fixed
$\eta_0$ the disk thickness increases as $\beta_{\rm r}$
increases, however, the ratio $h/R$ is more sensitive to the
variation of $\beta_{\rm r}$ for higher values of $\eta_0$.
Interestingly, as the magnetic diffusivity coefficient $\eta_0$
increases, the Toomre parameters $Q$ and $Q_{\rm M}$
significantly increase to values higher than unity.

The $\alpha$-dependence of the solutions is not strong as long as
$\beta_{\rm r}\approx |\beta_{\varphi}| \approx 1$. But we found
that the ratio $\beta$ and the Toomre parameter change because of
viscosity variations. Figure 3 (top, left) shows $\beta$ as a
function of $\beta_{\rm r}$. Clearly, as the viscosity parameter
$\alpha$ increases, the ratio $\beta$ increases. Also, the Toomre
parameter increases with the viscosity parameter. We found that
$\beta_{\varphi}$-dependence of the solutions is weak for
$|\beta_{\varphi}|\approx 1$. However, there is some changes on
the ratio $\beta$ as $\beta_{\varphi}$ varies. Figure 3 (bottom,
left) shows variation of $\beta$ as a function of $\beta_{\rm r}$
for $\beta_{\varphi}= -0.8$, $-1.0$ and $-1.2$. Evidently, this
ratio increases with $|\beta_{\varphi}|$. Also, by increasing
$|\beta_{\varphi}|$  the Toomre parameter increases, though it is
not very significant.

Bertin (1997) showed that the opening angle of self-regulated
nonmagnetized disk depends only on $Q$, independent of the
viscosity coefficient and the mass accretion rate. He argued that 
significantly large values of $Q$ would be undesirable because
they conflict with the thin disk approximation. But the disk
thickness in our magnetized case depends on the mass accretion
rate and the other input parameters as long as the beta factors
are of order unity. For larger values of $\beta_{\rm r}$ or
$\beta_{\varphi}$, the disk thickness may change because of the
magnetic fields.

\section{Conclusions}

We have obtained a self-similar solution for a self-gravitating,
magnetized, viscous disk.  The solution has constant rotational
and accretion velocities, independent of the radial distance.
Since we are considering an isothermal magnetized disk, the sound
speed is constant.

Our goal is to study the possible effects of the magnetic fields
on the steady state structure of self-gravitating, magnetized
disk, at least at the physical level. Although we have made some
simplifying assumptions in order to treat the problem
analytically, our self-similar solution shows magnetic fields can
really change typical behaviour of the physical quantities of a
self-gravitating disk. Not only the surface density of the disk
changes, but also the rotational and the radial velocities
significantly change because of the magnetic fields. It means any
realistic model for self-gravitating disk should consider the
possible effects of the magnetic fields. Of course, our
self-similar solutions are too simple to make any comparison with
observations. But, we think, one may relax self-similarity
assumption and solve the equations of the model numerically. In
doing so, our self-similar solutions can greatly facilitate
testing and interpretation of results. Then, we can calculate the
spectral energy distribution of such a self-gravitating
magnetized disk.

\begin{acknowledgements}
We are grateful to  the anonymous referee for the useful comments
which improved the manuscript. We also thank M. D. Smith, G. Bertin, R.
Pudritz and R. Lovelace for their very useful suggestions.
\end{acknowledgements}

\clearpage

\begin{figure}
\plotone{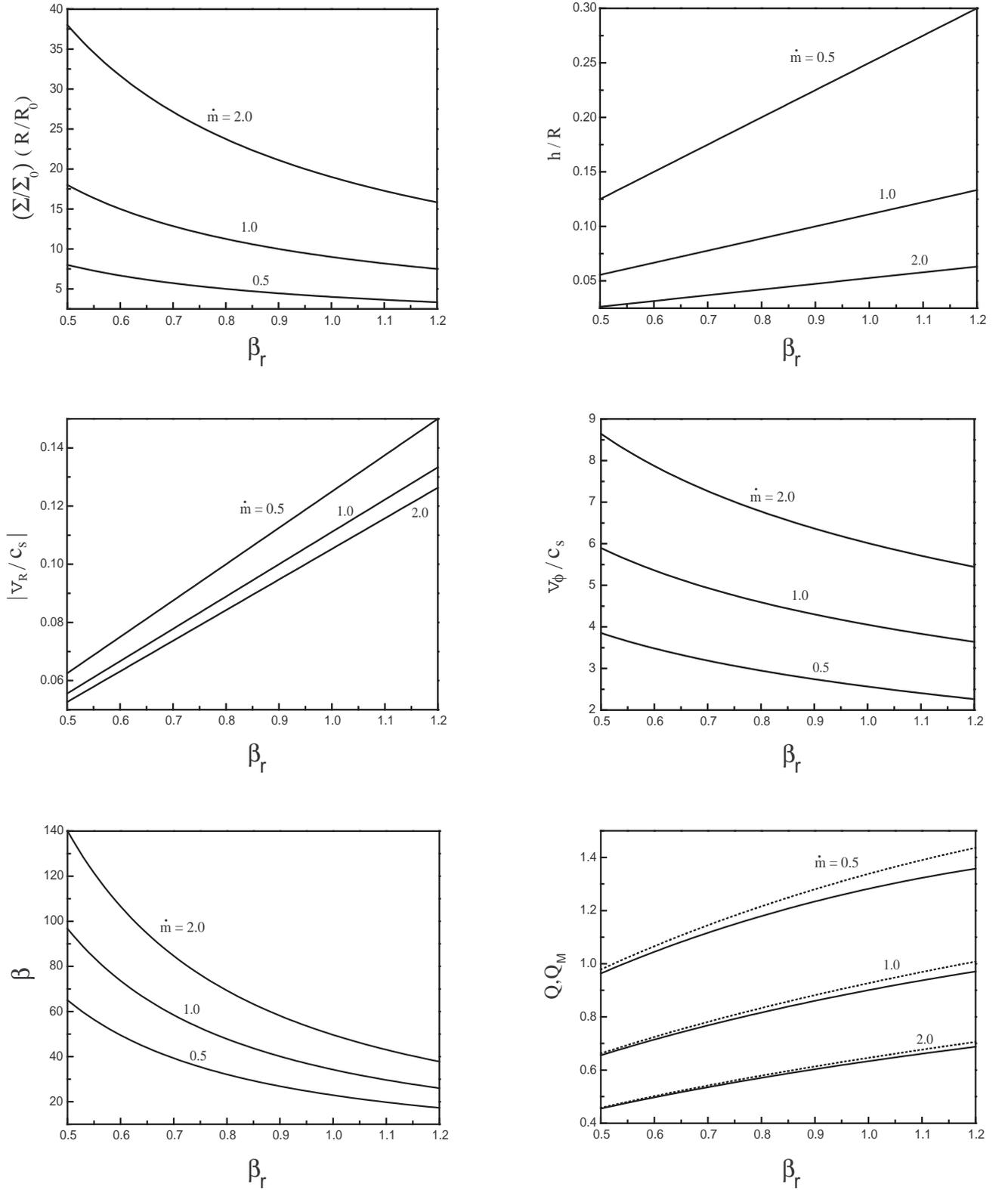} \caption{Typical behaviours of the surface
density (top, left), opening angle of the disk (top, right),
ratios of the radial and the rotational velocities to the sound
speed (middle), ratio of the thermal pressure to the magnetic
pressure at the surface (bottom, left) and the Toomre parameters
corresponding to $\alpha=0.1$, $\eta_{0}=0.1$,
$\beta_{\varphi}=-0.8$ and the other input parameters are
indicated on the plots. Magnetic Toomre parameter $Q_{\rm M}$ is
denoted by the dotted lines. These plots show the effects of the
mass accretion rate $\dot{m}$ on the structure of the disk.
}\label{fig:figure1}
\end{figure}
\begin{figure}
\plotone{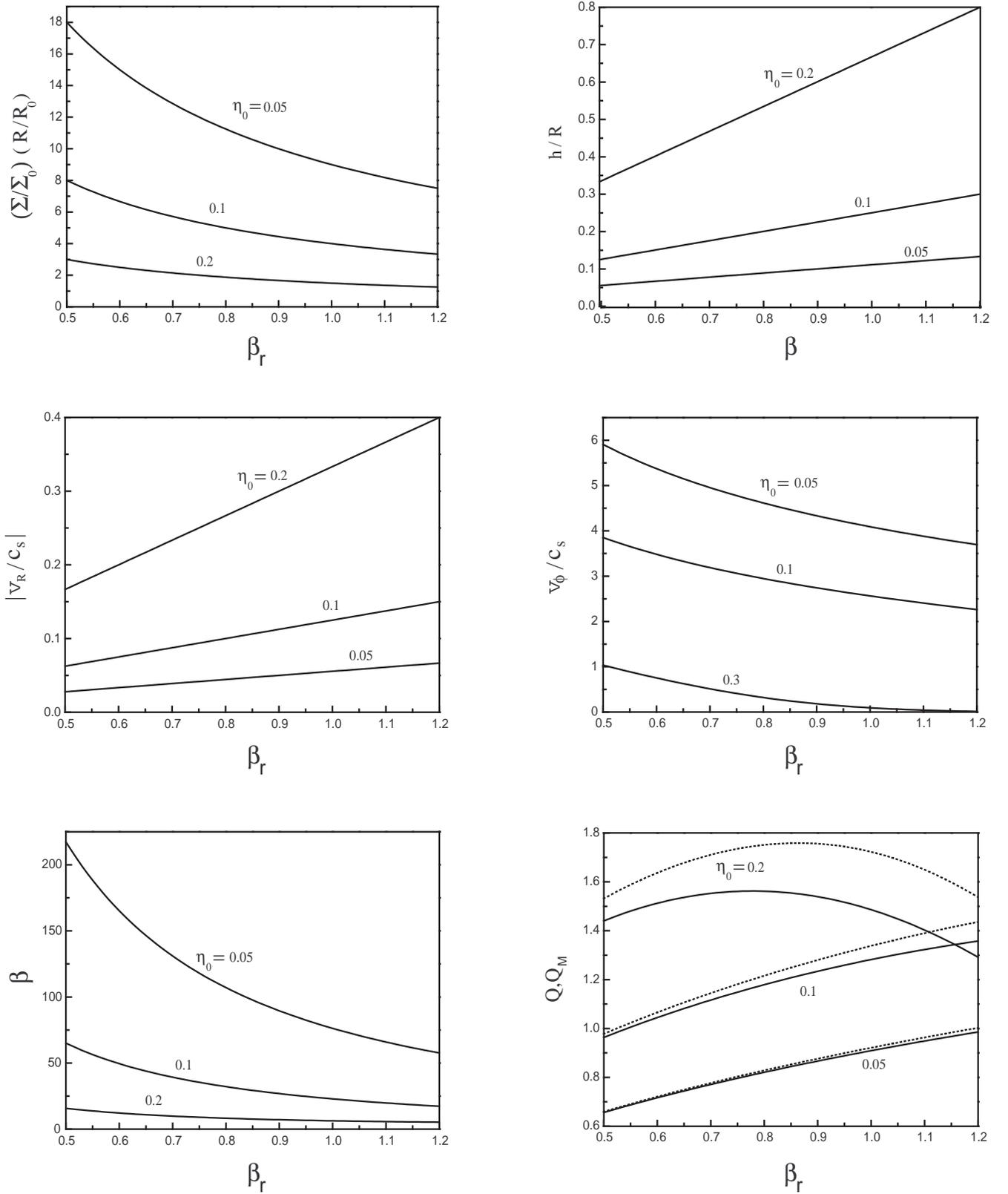}\caption{The same as Figure 1, but for
$\dot{m}=0.5$, $\alpha=0.1$, and $\beta_{\varphi}=-0.8$. These
plots show the effects of magnetic diffusivity coefficient
$\eta_0$ on the solutions. Also, the magnetic Toomre parameter is
denoted by dotted lines.}
\end{figure}
\begin{figure}
\plotone{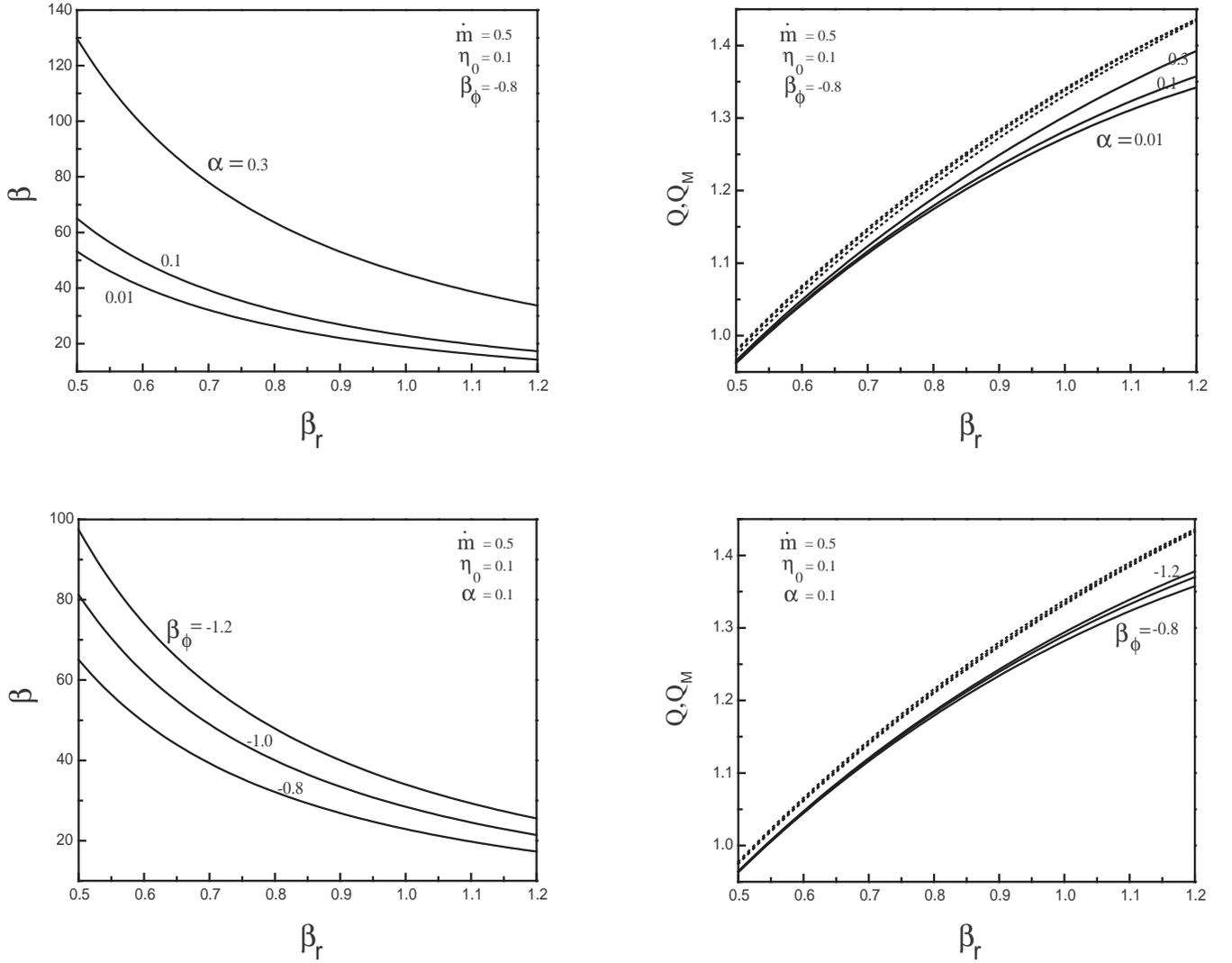}\caption{Top plots show the effects of the
viscosity parameter $\alpha$ on the structure of the disk. Among
the physical variables describing the disk, however, the ratio of
the thermal pressure to the magnetic pressure at the surface of
the disk $\beta$ and the Toomre parameters are more sensitive to
the variations of the viscosity coefficient. We have similar
behaviour when the parameter $\beta_{\varphi}$ varies, i.e. the
ratio $\beta$ and the Toomre parameters $Q$ and $Q_{\rm M}$ are
more sensitive to the variations of $\beta_{\varphi}$ keeping the
rest of the input parameters fixed (bottom plots).}
\end{figure}
\end{document}